\documentstyle[12pt]{article}
\def\beqar {\begin{eqnarray}}
\def\eeqar {\end{eqnarray}}
\def\beq {\begin{equation}}
\def\eeq {\end{equation}}
\def\F{{\cal F}}

\def\H{{\cal H}}

\def\S{{\cal S}}

\def\L{{\cal L}}

\def\M{{\cal M}}

\def\al{\alpha}
\def\bt{\beta}
\def\del{\delta}

\def\ga{\gamma}
\def\Ga{\Gamma}
\def\ka{\kappa}
\def\ep{\epsilon}
\def\la{\lambda}

\def\Om{\Omega}
\def\th{\theta}

\def\si{\sigma}

\def\d{\partial}

\def\hf{\frac{1}{2}}

\def\<{\langle}
\def\>{\rangle}
\def\Tr{{\rm Tr}}

\def\tr{{\rm tr}}

\def\dim{{\rm dim}}
\def\cp{{\bf CP}}
\begin{document}

\begin{titlepage}
\null\vspace{-62pt} \pagestyle{empty}
\begin{center}
%\rightline{} \rightline{CCNY-HEP-/05}
\vspace{1.0truein}

{\Large\bf Emergence of longitudinal 7-branes and fuzzy $S^4$ \\
\vspace{.3cm}
in compactification scenarios of M(atrix) theory} \\

%%%%%%%%%%%%%%%%%%%%%%%%%%%%%%%%%%%%%%%%%%%%%%%%%%%%%%%%%%
\vspace{.95in} Yasuhiro Abe\footnote{
Current address:
{\it Cereja Technology Co., Ltd. 3-1 Tsutaya-Bldg. 5F, Shimomiyabi-cho
Shinjuku-ku, Tokyo 162-0822, Japan}
({\tt abe@cereja.co.jp})
} \\
\vskip .12in {\it Physics Department\\ City College of the
CUNY\\
New York, NY 10031}\\
\vskip .07in {\tt abe@sci.ccny.cuny.edu}\\
\vspace{.95in}
%%%%%%%%%%%%%%%%%%%%%%%%%%%%%%%%%%%%%%%%%%%%%%%%%%%%%%%%%%%%
\centerline{\large\bf Abstract}
\end{center}
In M(atrix) theory, there exist membranes and longitudinal 5-branes
(L5-branes) as extended objects. Transverse components of these
brane solutions are known to be described by fuzzy $\cp^k$ ($k=1,2$),
where $k=1$ and $k=2$ correspond to spherical membranes and L5-branes of $\cp^2
\times S^1$ world-volume geometry, respectively.
In addition to these solutions, we here show the
existence of L7-branes of $\cp^3 \times S^1$ geometry, introducing extra
potentials to the M(atrix) theory Lagrangian.
As in the cases of $k=1,2$, the L7-branes (corresponding to $k=3$)
also break the supersymmetries of M(atrix) theory.
The extra potentials are introduced
such that the energy of a static L7-brane solution becomes finite in the
large $N$ limit where $N$ represents the matrix dimension of fuzzy $\cp^3$.
As a consequence, fluctuations from the L7-branes are suppressed,
which effectively describes compactification of M(atrix) theory down to 7 dimensions.
We show that one of the extra potentials can be considered as a matrix-valued 7-form.
The presence of the 7-form in turn supports a possibility of Freund-Rubin type
compactification. This suggests that
our modification of M(atrix) theory can also lead to a physically interesting
matrix model in four dimensions.
In hope of such a possibility, we further consider compactification of
M(atrix) theory down to fuzzy $S^4$ which can be defined in terms of fuzzy $\cp^3$.
Along the way, we also find a new L5-brane solution to
M(atrix) theory which has purely spherical geometry in the transverse directions.

\end{titlepage}
%%%%%%%%%%%%%%%%%%%%%%%%%%%%%%%%%%%%%%%%%%%%%%%%%%%%%%%%%%%%%
\pagestyle{plain} \setcounter{page}{2} %\baselineskip =14pt

\section{Introduction}

There has been extensive interest in the matrix model of M-theory or
the so-called M(atrix) theory since its proposal
by Banks, Fischler, Shenker and Susskind (BFSS) \cite{BFSS}.
For a review of M(atrix) theory, one may refer to \cite{Review1}.
In M(atrix) theory, 9 dimensions out of
11 are described by $(N\times N)$-matrices, while the other
dimensions correspond to light-front coordinates. This structure arises
as a natural extension of matrix regularization of bosonic membranes in
light-front gauge.
The ordinary time component and the extra spatial
direction, the so-called longitudinal one, emerge from the light-front
coordinates in M(atrix) theory. The longitudinal coordinate is
considered to be toroidally compactified with a radius $R$.
In this way, the theory can be understood in 10 dimensions.
This is in accordance with one of the features of M-theory, {\it i.e.},
as a strongly coupled limit of type {\rm II}A string theory,
since the radius $R$ can be related to the string coupling constant $g$
by $R=g l_s$ where $l_s$ is the string length scale.
From a 11-dimensional viewpoint, one can consider
certain objects which contain a longitudinal momentum $N/R$ as a
Kaluza-Klein mode.
Partly from these observations it has been
conjectured that the large $N$ limit of M(atrix) theory should describe
M-theory in the large longitudinal momentum limit or in the so-called
infinite momentum frame (IMF).
This BFSS conjecture has been confirmed
in various calculations, especially in regard to perturbative
calculations of graviton interactions (see, e.g., \cite{KT2,OY}),
capturing another feature of M-theory, {\it i.e.}, emergence of 11-dimensional
supergravity in the low energy limit.
There also exits a related matrix
model by Ishibashi, Kawai, Kitazawa and Tsuchiya (IKKT) \cite{IKKT}
which corresponds to type IIB string theory.
This IKKT model has been investigated with a lot of attention as well.
For a review of this model, one may refer to \cite{Review2}.

Besides gravitons, M(atrix) theory further contains extended and charged
objects, namely, memberanes and 5-branes.
The membrane in matrix context appeared originally in the quantization
of the supermembrane many years ago by de Wit, Hoppe and Nicolai \cite{deWit}.
Membranes of spherical geometry in M(atrix) theory have been obtained in \cite{KT1,Rey}.
As regards 5-branes, they were obtained as longitudinal 5-branes or
L5-branes \cite{BD,GRT,BSS}.
The L5-branes are named after the property that one of their five dimensions
coincides with the longitudinal direction in M(atrix) theory.
One may think of the existence of transverse 5-branes as opposed to L5-branes
but it turns out that there are no classically conserved charges
corresponding to the transverse 5-branes.
Thus it is generally believed that the L5-branes are the only
relevant 5-branes in M(atrix) theory at least in the classical level \cite{Mald1}.
In a modified M(atrix) theory, {\it i.e.}, the so-called plane wave matrix
theory \cite{BMN}, the existence of transverse 5-branes is discussed at
a quantum level \cite{Mald1}.
L5-branes with spherical geometry in the
transverse directions have been proposed in \cite{CLT}.
Although this spherical L5-brane captures many properties of M-theory, it is as
yet unclear how to include matrix fluctuations contrary to the case of
spherical membranes.
The only other L5-brane that is known so far is an L5-brane with $\cp^2$ geometry
in the transverse directions \cite{Nair1}.
Matrix configuration of this L5-brane is relevant to that of the fuzzy $\cp^2$ \cite{Bal1}.

Fuzzy spaces are one of the realizations of noncommutative geometry
\cite{Connes} in terms of $(N\times N)$-matrices, hence, those extended
objects in M(atrix) theory are possibly described by the fuzzy spaces as
far as the transverse directions are concerned.
Following this idea, in the present paper we shall consider
the fuzzy complex projective spaces $\cp^k$ ($k=1,2,\cdots$)
as ans\"{a}tze to the extended objects or the brane solutions in M(atrix) theory.
This approach towards a solution to
M(atrix) theory was originally pursued by Nair and Randjbar-Daemi in
\cite{Nair1} which, among the other known brane solutions, revealed the
existence of the L5-brane of $\cp^2 \times S^1$ geometry. The fuzzy
$\cp^k = SU(k+1)/U(k)$ can generally be constructed in terms of matrix
representations of the algebra of $SU(k+1)$ in the
$(n,0)$-representation (or the totally symmetric representation of rank
$n$) under a certain set of algebraic constraints \cite{Bal2}. This fact
makes it relatively straightforward to include transverse fluctuations
of branes with $\cp^2$ (or $\cp^k$) geometry in comparison with the case
of the spherical L5-brane. This point is one of the advantages to
consider the fuzzy $\cp^k$ as ans\"{a}tze for the brane solutions. Note
that fluctuations of branes are described by gauge fields on
noncommutative geometry. This means that the dynamics of the extended
objects in M(atrix) theory can be governed by gauge theories on fuzzy
spaces. (For a general review of noncommutative field theory, see, for
instance, \cite{NC}. For a recent review of fuzzy spaces in relation to
the M(atrix) theory as well as to the quantum Hall effect, see
\cite{Nair2}.)

From a perspective of type IIA string theory, the gravitons, membranes
and L5-branes of M-theory are respectively relevant to D0, D2 and D4
brane solutions. Type IIA string theory also contains a D6 brane. The D6
brane is known to be a Kaluza-Klein magnetic monopole of 11-dimensional
supergravity compactified on a circle and is considered to be irrelevant
as a brane solution in M(atrix) theory. Naively, however, since D6
branes are Hodge dual to D0 branes in the same sense that D2 and D4
branes are dual to each other, we would expect the existence of
L7-branes in M(atrix) theory. It is important to note that fuzzy spaces
can be constructed only for compact spaces. If we parametrize branes by
fuzzy spaces, the transverse directions are also all compactified in the
large $N$ limit. As far as the capture of a Kaluza-Klein mode in the
scale of $N/R$ is concerned, one cannot distinguish the longitudinal
direction from the transverse ones. The gravitons or the corresponding
D0 branes of M-theory would possibly live on the transverse directions
in this case. Thus we may expect the existence of L7-branes as a Hodge
dual description of such gravitons in an M-theory perspective.
Construction of L7-branes (or transverse D6-branes) has been suggested
in \cite{BSS,Taylor1}, however, such extended objects have not been
obtained in the matrix model. Besides the fact that no L7-brane charges
appear in the supersymmetry algebra of M(atrix) theory, there is a
crucial obstruction to the construction of L7-brane, that is, as shown
by Banks, Seiberg and Shenker \cite{BSS}, the L7-brane states have an
infinite energy in the large $N$ limit, where the energy of the state is
interpreted as an energy density in the transverse directions. Indeed,
as we shall discuss in the next section, an L7-brane of $\cp^3 \times
S^1$ geometry leads to an infinite energy in the large $N$ limit and,
hence, one cannot make sense of the theory with such an L7-brane.

In order to obtain an L7-brane as a solution to M(atrix) theory, it
would be necessary to introduce extra potentials or fluxes to the
M(atrix) theory Lagrangian such that the brane system has a finite
energy as $N \rightarrow \infty$. Since M(atrix) theory is defined on a
flat space background, such an additional term suggests the description
of the theory in a nontrivial background. The most notable modification
of the M(atrix) theory Lagrangian would be the one given by Berenstein,
Maldacena and Nastase (BMN) to describe the theory in the maximally
supersymmetric parallel-plane (pp) wave background \cite{BMN}.
There has been a number of papers on this BMN matrix model of M-theory.
(For some earlier papers, see \cite{works1}.)
Another important approach to the modification of BFSS M(atrix) theory is to
introduce a Ramond-Ramond (RR) field strength as a background such that
it couples to brane solutions.
Specifically, one may have a RR 4-form as an extra potential
from a IIA string theory viewpoint.
As shown by Myers \cite{Myers}, the matrix equation of motion with
this RR flux allows fuzzy $S^2$ ($= \cp^1$) as a static solution,
meaning that the corresponding IIA theory has a spherical D2-brane solution.
The RR field strength is associated with a charge of this D2 brane.
The modified equation of motion also allows a diagonal matrix configuration
as a solution which corresponds to $N$ D0-branes, with $N$ being the dimension of matrices.
One may interpret these solutions as bound states
of a spherical D2-brane and $N$ D0-branes.
From a D0-brane perspective, the RR field strength is also associated with a D0-brane charge.
Thus the extra RR flux gives rise to a D-brane analog of a dielectic effect, which is
known as Myers effect.
A different type of flux, {\it i.e.}, a RR 5-form which
produces bound states of $N$ D1-branes and a D5-brane with $\cp^2$
geometry has been proposed by Alexanian, Balachandran and Silva
\cite{Bal3} to describe a generalized version of Myers effect from a
viewpoint of IIB string theory.
From a perspective of M(atrix) theory, the D5 brane corresponds to
the L5-brane of $\cp^2 \times S^1$ geometry.
In this paper, we consider further generalization along these lines of developments.
Namely, we consider a general form for all possible extra potentials that allows
fuzzy $\cp^k$ as brane solutions or solutions of modified matrix equations of motion.
We find several such potentials for $k \le 3$.

The extra potentials we shall introduce in the consideration of a
possible L7 brane solution to M(atrix) theory are relevant to fluxes on
a curved space of $(\cp^3 \times S^1) \times \M_4$ where $\M_4$ is an
arbitrary four-dimensional manifold. We shall show that one of the
potentials can be interpreted as a 7-form flux in M(atrix) theory.
According to Freund and Rubin \cite{FR}, existence of a 7-form in 11
dimensional (bosonic) theories implies compactification of 7 or 4
space-like dimensions. The existence of the 7-form in M(atrix) theory is
interesting in a sense that it would lead to a matrix version of
Freund-Rubin type compactification.
This means that the introduction of the 7-form can also lead to
a physically interesting matrix model in four dimensions.
In hope of such a possibility, we also consider
compactification of M(atrix) theory down to fuzzy $S^4$
which can be defined in terms of fuzzy $\cp^3$ \cite{Abe1}.

The plan of the rest of this paper is as follows. In the next section,
following Nair and Randjbar-Daemi \cite{Nair1}, we show that the fuzzy
$\cp^k$ ($k\le 4$) provide solutions to bosonic matrix configurations in
the M(atrix) theory Lagrangian. Along the way we briefly present
definitions and properties of fuzzy $\cp^k$. We further discuss the
energy scales of the solutions and see that the energy becomes finite in
the large $N$ limit only in the cases of $k=1,2$, corresponding to the
membrane and the L5-brane solutions in M(atrix) theory. In section 3, we
examine supersymmetry of the brane solutions for $k \le 3$. We make a
group theoretic analysis to show that those brane solutions break the
supersymetries in M(atrix) theory. Our discussion is closely related to
the previous analysis \cite{Nair1} in the case of $k=2$. In section 4,
we introduce extra potentials to the M(atrix) theory Lagrangian which
are suitable for the fuzzy $\cp^k$ brane solutions. We consider the
effects of two particular potentials to the theory. These effects can be
considered as generalized Myers effects. We find a suitable form of
potentials for the emergence of static L7-brane solutions, such that the
potentials lead to finite L7-brane energies in the large $N$ limit.
Section 5 is devoted to the discussion on possible compactification
models in non-supersymmetric M(atrix) theory. We show that one of the
extra potentials introduced for the presence of L7-branes can be
interpreted as a matrix-valued or {\it fuzzy} 7-form in M(atrix) theory.
Using the idea of Freund-Rubin type compactification,
this suggests the compactification down to 7 or 4 dimensions.
The compactification model down to 4 dimension is physically the more interesting
and we consider, as a speculative model of it, a compactified
matrix model on fuzzy $S^4$.
Lastly, we present brief conclusions.

%%%%%%%%%%%%%%%%%%%%%%%%%%%%%%%%%%%%%%%%%%%%%%%%%%%%%%%%%%%%%%%%%%%%%
\section{Fuzzy $\cp^k$ as brane solutions to M(atrix) theory}

The M(atrix) theory Lagrangian can be expressed as
\beq
    \L = \Tr \left(
    \frac{1}{2R} {\dot X}_{I}^{2} + \frac{R}{4} [X_I , X_J ]^2 +
    \th^{T} {\dot \th} + i R \th^{T} \Ga_I [X_I , \th] \right)
    \label{c1}
\eeq
where $X_I$ ($I=1,2,\cdots, 9$) are hermitian $N\times N$ matrices,
$\th$ denotes a 16-component spinor of $SO(9)$ represented by $N\times
N$ Grassmann-valued matrices, and $\Ga_I$ are the $SO(9)$ gamma matrices
in the 16-dimensional representation. The Hamiltonian of the theory is
given by
\beq
    \H = \Tr \left( \frac{R}{2} P_I P_I - \frac{R}{4} [X_I, X_J]^2
    -i R \th^{T} \Ga_I [X_I, \th ] \right)
    \label{c2}
\eeq
where $P_I$
is the canonical conjugate to $X_I$; $\frac{\d \L}{\d \dot{X}_I}$. As
discussed in the introduction, we will be only interested in those
energy states that have finite energy in the limit of the large
longitudinal momentum $N/R$. Since the Hamiltonian (\ref{c2}) leads to
an infinite energy state in the limit of $R \rightarrow \infty$, we will
consider the large $N$ limit with a large, but fixed value for $R$. With
this limit understood, the theory is defined by (\ref{c1}) or (\ref{c2})
with a subsidiary Gauss law constraint
\beq
    [ X_I , \dot{X}_I ] - [ \th , \th^{T} ] = 0 \, .
    \label{c3}
\eeq
In this section, we shall consider the
bosonic part of the theory, setting the $\th$'s to be zero. The relevant
equations of motion for $X_I$ are given by
\beq
    \frac{1}{R} {\ddot X}_I - R [ X_J , [X_I , X_J]] = 0
    \label{c4}
\eeq
with a subsidiary constraint
\beq
    [X_I , \dot{X}_I] = 0 \, .
    \label{c5}
\eeq

We shall look for solutions to these equations, taking the
following ans\"{a}tze
\beq
    X_I  =  \left\{
    \begin{array}{ll}
      r(t) Q_i & \mbox{for $I=i=1,2,\cdots, 2k$}\\
      0  & \mbox{for $I=2k+1, \cdots, 9$}
    \end{array} \right.
    \label{c6}
\eeq
where $Q_i$ denote the local coordinates of fuzzy $\cp^k =
SU(k+1)/U(k)$ ($k=1,2,\cdots$). Since $X_I$ are defined for
$I=1,2,\cdots, 9$, the ans\"{a}tze are only valid for $k \le 4$.

\vspace{0.4cm}
 \noindent \emph{2.1 \underline{Construction of fuzzy $\cp^k$: a review}}
\vspace{0.2cm}

The fuzzy $\cp^k$ can be constructed in terms of certain matrix
generators of $SU(k+1)$ as embedded in ${\bf R}^{k^2 + 2k}$ under
a set of algebraic constraints. Here we shall briefly review such
a construction, following a description in \cite{Abe1}. Let $L_A$
be $N^{(k)} \times N^{(k)}$-matrix representations of $SU(k+1)$
generators in the $(n,0)$-representation (or the totally symmetric
representation of rank $n$). The coordinates of fuzzy $\cp^k$ as
embedded in ${\bf R}^{k^2 + 2k}$ are defined by
\beq
    Q_A = \frac{L_A}{\sqrt{C^{(k)}_2}}
    \label{a1}
\eeq
with two constraints
\beqar
    Q_A Q_A &=& {\bf 1}  \label{a2} \\
    d_{ABC} Q_{A}  Q_{B} &=& c_{k,n} Q_C
    \label{a3}
\eeqar
where
${\bf 1}$ is the $N^{(k)} \times N^{(k)}$ identity matrix,
$d_{ABC}$ is the totally symmetric symbol of $SU(k+1)$,
$C^{(k)}_2$ is the quadratic Casimir of $SU(k+1)$ in the
$(n,0)$-representation
\beq
    C^{(k)}_2 = \frac{n k (n+k+1)}{2 (k+1)}
    \label{a4}
\eeq
and $N^{(k)}$ is the dimension of $SU(k+1)$
in the $(n,0)$-representation
\beq
    N^{(k)} = \dim (n,0) =
    \frac{(n+k)!}{k!~n!} \, .
    \label{a5}
\eeq
The coefficient $c_{k,n}$
in (\ref{a3}) is given by
\beq
    c_{k,n} =
    \frac{(k-1)}{\sqrt{C_2^{(k)}}} \left( \frac{n}{k+1} + \hf
    \right) \, .
    \label{a6}
\eeq
For $k \ll n$, we have
\beq
    c_{k,n} \, \longrightarrow \, c_{k} = \sqrt{\frac{2}{k(k+1)}} (k-1)
    \label{a7}
\eeq
and this leads to the constraints for the
coordinates $q_A$ of commutative ${\bf CP}^k$
\beqar
    q_A q_A &=& 1 \, ,
    \label{a8} \\
    d_{ABC} q_{A} q_{B} & = & c_{k} q_C \, .
    \label{a9}
\eeqar
The second constraint (\ref{a9}) restricts the number of coordinates
to be $2k$ out of $k^2+2k$. For example, in the case of
$\cp^2=SU(3)/U(2)$ this constraint around the pole of $A=8$
becomes $d_{8BC}q_8 q_B = \frac{1}{\sqrt{3}} q_C$. Normalizing the
8-coordinate to be $q_8 = -2$, we find that the indices of the
coordinates are restricted to 4, 5, 6, and 7 with the conventional
choice of the generators of $SU(3)$. Similarly, under the
constraints (\ref{a3}), the coordinates of fuzzy $\cp^k$ are
effectively expressed by the local coordinates $Q_i$
($i=1,2,\cdots, 2k$) rather than the global ones $Q_A$
($A=1,2,\cdots, k^2+2k$).

\vspace{0.4cm}
 \noindent \emph{2.2 \underline{Local coordinates of fuzzy $\cp^k$}}
\vspace{0.2cm}

We now consider the commutation relations of the fuzzy $\cp^k$ coordinates $Q_i$.
By construction, they are embedded in the $SU(k+1)$ algebra.
We first split the generators
$L_A$ of $SU(k+1)$ into $L_i \in \underline{SU(k+1)}-\underline{U(k)}$
and $L_\al \in \underline{U(k)}$, where $\underline{G}$ denotes the Lie
algebra of group $G$.
The indices $i=1,2, \cdots, 2k$ are then relevant to the
$\cp^k$ of our interest, while the indices $\al = 1,2,\cdots, k^2$ correspond
to a $U(k)$ subgroup of $SU(k+1)$.
The $SU(k+1)$ algebra,
$[ L_A, L_B]= if_{ABC}L_C$ with the structure constant $f_{ABC}$,
is then expressed by the following set of commutation relations
\beqar
\left[Q_i, Q_j \right] &=& i ~
\frac{c_{ij\al}}{\sqrt{C^{(k)}_2}}~Q_\al \label{c7}\\
\left[Q_{\al}, Q_{\bt} \right] &=& i~
\frac{f_{\al\bt\ga}}{\sqrt{C^{(k)}_2}}~Q_{\ga} \label{c8}\\
\left[Q_{\al}, Q_i \right] &=& i ~
\frac{f_{\al ij}}{\sqrt{C^{(k)}_2}}~Q_j \label{c9}
\eeqar
where we use $Q_A= L_A/{\sqrt{C^{(k)}_2}}$ and denote $f_{ij\al}$
by $c_{ij\al}$ to indicate that it is relevant to the commutators
of $Q_i$'s. $f_{\al\bt\ga}$ is essentially the structure constant
of $SU(k)$ since the $U(1)$ part of the $U(k)$ algebra can be
chosen such that it commutes with the rest of the algebra. We can
calculate $c_{\al ij} c_{\bt ij}$ as
\beq
c_{\al ij} \, c_{\bt ij}
\, = \, f_{\al AB} f_{\bt AB} -f_{\al \ga \del} f_{\bt \ga \del}
\,=\, \del_{\al\bt}
\label{c10}
\eeq
by use of the relations
$f_{\al AB} f_{\bt AB}= (k+1)\del_{\al\bt}$ and $f_{\al \ga \del}
f_{\bt \ga \del} = k \del_{\al\bt}$. Notice that the result
(\ref{c10}) restricts possible choices of the $\cp^k$ indices
$(i,j)$. For example, in the case of $k=2$ we have $(i,j) = (4,5),
(6,7)$ with the conventional choice of the structure constant
$f_{ABC}$ of $SU(3)$. Similarly, in the case of $k=3$ we have
$(i,j)=(9,10),(11,12),(13,14)$. Under such restrictions, we can
also calculate $c_{ij\al} f_{j\al k}$ as
\beq
c_{ij\al} \, f_{j\al k} \, = \, c_{ij\al} \, c_{kj\al} \, = \, \del_{ik} \, .
\label{c11}
\eeq
In what follows, we shall use the symbol $c_{ij\al}$
rather than $f_{ij \al}$ to indicate that we are interested in
this peculiar subset of the $SU(k+1)$ algebra.

We can also classify the totally symmetric symbol $d_{ABC}$ as follows.
\beq
    d_{ABC} = \left\{ \begin{array}{ll}
    d_{ij \al} & \\
    d_{\al \bt \ga} & \\
    0 & \mbox{otherwise}
    \end{array}\right.
    \label{c11-1}
\eeq
Notice that symbols such as $d_{\al \bt i}$ and $d_{ijk}$ do vanish.
In relation to the construction of $\cp^k$, it is useful to know
the fact that the symbol $d_{ii\al}$, a subset of $d_{ij \al}$, is expressed
as $d_{ii \al_{k^2 + 2k}}$ and is identical regardless the index $i$.
Here the index $i$ is relevant to a local coordinate of $\cp^k$
and the index $\al_{k^2 + 2k}$ is a hypercharge-like index in
a conventional choice of $SU(k+1)$ generators.

The normalization of $Q_A$'s is taken as (\ref{a2}). Thus
traces of matrix products are expressed as
\beqar
    \Tr(Q_{A} Q_{B}) &=& \frac{N^{(k)}}{k^2 +2k} \del_{AB} \, , \label{c12} \\
    \Tr(Q_{i} Q_{i}) &=& \frac{2k}{k^2 +2k} N^{(k)}  \, , \label{c13} \\
    \Tr(Q_{\al} Q_{\al}) &=& \frac{k^2}{k^2 +2k} N^{(k)} \, .  \label{c14}
\eeqar
These relations are also useful in later calculations.

\vspace{0.4cm}
 \noindent \emph{2.3 \underline{Fuzzy $\cp^k$ solutions to M(atrix) theory}}
\vspace{0.2cm}

Using (\ref{c7})-(\ref{c11}), we can easily find
that $[Q_j , [Q_i , Q_j]]= - Q_i / C^{(k)}_2$.
Thus, with  the fuzzy $\cp^k$ ans\"{a}tze (\ref{c6}),
the equation of motion (\ref{c4}) becomes
\beq
    \left( \frac{\ddot{r}}{R} + \frac{R}{C^{(k)}_2} r^3 \right) Q_{i} = 0 \, .
    \label{c15}
\eeq
This means that the equation of motion is reduced to
an ordinary differential equation of $r(t)$.
Notice that the subsidiary constraint (\ref{c5}) is also satisfied with
the ans\"{a}tze (\ref{c6}).
The equation of motion therefore reduces to
\beq
    \ddot{r} + \frac{R^2}{C^{(k)}_2} r^3 = 0 \, .
\label{c16}
\eeq
A general solution to this equation is written as
\beq
    r(t) \, = \,  A \, {\rm cn} \left( \al(t-t_0);\ka^2 =\hf \right)
    \label{c17}
\eeq
where $\al=\sqrt{R^2/{C^{(k)}_2}}$ and ${\rm cn}(u; \ka)={\rm
cn}(u)$ is one of the Jacobi elliptic functions, with $\ka$ ($0
\le \ka \le 1$) being the elliptic modulus. $A$ and $t_0$ are the
constants determined by the initial conditions. Using the following formula
\beqar
    \frac{d}{du}{\rm cn} (u; \ka)&=& - \,{\rm sn}(u; \ka) \,
    {\rm dn}(u;\ka) \nonumber\\ &=& - u + \frac{1+4\ka^2}{3!}u^3 - \cdots \, ,
    \label{c18}
\eeqar
we can express $\dot{r}$ as
\beq
\dot{r} \, = \, -A \al \, {\rm sn}(\al (t-t_0)) {\rm dn}(\al (t-t_0)) \,.
\label{c19}
\eeq
In the limit of large $N$ (or $n$), $\dot{r}$ is
suppressed by $\dot{r} \sim 1/n^2$. Thus the solution (\ref{c17})
corresponds to a static solution in the large $N$ limit.

Evaluated on the fuzzy $\cp^k$, the potential energy of M(atrix) theory is
calculated as
\beqar
    V (rQ) &=&  - \Tr \left( \frac{R}{4} [ r Q_{i} , r Q_{j} ]^2 \right)
    \nonumber \\
    &=& \frac{R r^4}{4 C_{2}^{(k)}} \Tr ( Q_{\al} Q_{\al} ) \nonumber \\
    &=& \frac{k^2}{k^2 + 2k }\frac{R r^4}{4 C_{2}^{(k)}} N^{(k)}
    \, \sim \, n^{k-2} R r^4 \, .
    \label{c20}
\eeqar
From this result we can easily tell that for
$k=1,2$ we have finite energy states in the large $N$ limit. These
states respectively correspond to the spherical membrane and the
L5-brane of $\cp^2$ geometry in M(atrix) theory. By contrast, for
$k=3,4$ we have infinite energy states. Thus, although these may
possibly correspond to L7 and L9 brane solutions, they are
ill-defined and we {\it usually} do not consider such solutions in
M(atrix) theory.
The main purpose of the present paper is to show that we can have L7 brane
solutions by introducing extra potentials to the M(atrix) theory
Lagrangian (\ref{c1}).
Notice that in this paper we shall not consider the
case of $k=4$ or a 9-brane solution to M(atrix) theory.
The 9-branes are supposed to correspond to ``ends of the world'' which
describe gauge dynamics of the 9-dimensional boundary of M-theory.
Thus these are in general considered irrelevant as brane solutions to the theory.

%%%%%%%%%%%%%%%%%%%%%%%%%%%%%%%%%%%%%%%%%%%%%%%%%%%%%%%%%%%%%
\section{Supersymmetry breaking}

In this section, we examine supersymmetry of the fuzzy $\cp^k$
brane solutions in M(atrix) theory for $k \le 3$.
As in the previous section, we make an analysis, following
the argument of Nair and Randjbar-Daemi in \cite{Nair1}.

We have set the fermionic matrix variables $\th$ to be zero.
Now we consider the supersymmetry transformations of
the brane solutions in M(atrix) theory. The supersymmetric
variation of $\th$ is given by
\beq
    \del \th_r = \hf \left(
    \dot{X}_I (\Ga_{I})_{rs} + [X_I , X_J] (\Ga_{IJ})_{rs} \right)
    \ep_s + \del_{rs}\xi_s
    \label{s1}
\eeq
where $\ep$ and $\xi$ are
16-component spinors of $SO(9)$ represented by $N \times N$
matrices ($r,s =1,2,\cdots,16$) and $\Ga_I$'s are the
corresponding gamma matrices as before. $\Ga_{IJ}$ are defined by
$\Ga_{IJ}= \hf [\Ga_I, \Ga_J]$.
With the fuzzy $\cp^k$ ans\"{a}tze (\ref{c6}), the equation (\ref{s1}) reduces to
\beq
    \del \th_r = \hf \left( \dot{r} Q_i
    (\ga_i)_{rs} +  r^2 \frac{i c_{ij\al}}{\sqrt{C^{(k)}_2}}\, Q_\al
    (\ga_{ij})_{rs} \right) \ep_s + \del_{rs} \xi_s
    \label{s2}
\eeq
where $\ga_i$'s are the gamma matrices of $SO(2k)$ under the
decomposition of $SO(9) \rightarrow SO(2k) \times SO(9-2k)$.
Accordingly, we here set $i=1,2,\cdots, 2k$ and $r,s=1,2,\cdots,
2^k$. For the static solution we make $\dot{r} \sim n^{-2}$
vanish. Indeed, if $\del \th \sim n^{-2}$, we have $\Tr (\del
\th^{T} \del \dot{\th}) \sim N^{(k)}n^{-4} \sim n^{k-4}$ and, for
$k=1,2$ and $3$, this term vanishes in the large $N$ limit. The
other term $\Tr (i R \del \th^{T} \Ga_I [X_I , \del\th])$ in the
Lagrangian vanishes similarly. Thus, for static solutions, the
condition $\del \th =0$ is satisfied when $c_{ij\al} Q_\al
\ga_{ij}$ becomes a ${\bf c}$-number in the $SO(2k)$ subspace of
$SO(9)$ such that the $\ep$-term can be canceled by $\xi$ in
(\ref{s2}). In what follows, we examine this 
Bogomol'nyi-Prasad-Sommerfield (BPS)-like condition
for $k=1,2,3$.

It is known that the spherical membrane solution breaks all
supersymmetries. Let us rephrase this fact by examining the BPS
condition ($\del \th =0$) for $k=1$. The 2-dimensional gamma
matrices are given by $\ga_1 = \si_1$ and $\ga_2 = \si_2$, where
$\si_i$ is the ($2\times 2$)-Pauli matrices. The factor
$c_{ij\al}Q_\al \ga_{ij}$ becomes proportional to $Q_3 \si_3$
where $Q_3$ is an $N^{(1)} \times N^{(1)}$ matrix representing the
$U(1)$ part of the $SU(2)$ generators in the spin-$n/2$
representation. Now the factor $\si_3$ is not obviously
proportional to identity in the $SO(2)$ subspace of $SO(9)$.
Thus we can conclude that the BPS condition is broken.

For $k=2$, we can apply the same analysis to the factor of
$c_{ij\al}Q_\al \ga_{ij}$. We use the conventional choice for the
structure constant of $SU(3)$ where the group elements are defined
by $g= \exp (i\th^a \frac{\la^a}{2})$ with the Gell-Mann matrices
$\la^a$ ($a=1,2,\cdots, 8)$. As discussed earlier, with this
convention the set of $(i,j)$ is restricted to $(i,j)=(4,5)$ or
$(6,7)$. The relevant $c_{ij\al}$'s are given by $c_{453}=1/2$,
$c_{458}=\sqrt{3}/2$, $c_{673}=-1/2$ and $c_{678}=\sqrt{3}/2$.
Introducing the usual 4-dimensional gamma matrices $\ga_i$
($i=4,5,6,7$)
\beq
    \ga_4 = \left(%
    \begin{array}{cc}
      0 & 1 \\
      1 & 0 \\
    \end{array}%
    \right), ~ \ga_5 =\left(%
    \begin{array}{cc}
     0 & -i\si_1 \\
      i\si_1 & 0 \\
    \end{array}%
    \right), ~ \ga_6=\left(%
    \begin{array}{cc}
      0 & -i\si_2 \\
      i\si_2 & 0 \\
    \end{array}%
    \right),~ \ga_7 =\left(%
    \begin{array}{cc}
     0 & -i\si_3 \\
      i\si_3 & 0 \\
    \end{array}%
    \right),
    \label{s3}
\eeq
 we can calculate the factor of interest as
\beqar
    c_{45\al} Q_\al \ga_{45}
    &\sim &
     \left( Q_3+ \sqrt{3} Q_8 \right) \left(%
    \begin{array}{cc}
      i\si_1 & 0 \\
      0 & -i\si_1 \\
    \end{array}%
    \right) \nonumber\\
    c_{67\al} Q_\al \ga_{67}
    &\sim &
     \left( -Q_3+ \sqrt{3} Q_8 \right) \left(%
    \begin{array}{cc}
      i\si_1 & 0 \\
      0 & i\si_1 \\
    \end{array}%
    \right)
    \label{s4}
\eeqar
where $Q_3$ and $Q_8$ are $N^{(2)} \times
N^{(2)}$ matrices representing diagonal parts of $SU(3)$ algebra
in the totally symmetric representation $(n,0)$. In either case,
it is impossible to make the factor $c_{ij\al}Q_\al \ga_{ij}$ be
proportional to identity or zero in terms of the $(4\times
4)$-matrix which corresponds to $\ga_i$'s. This indicates that the
brane solution corresponding to $k=2$ breaks the supersymmetries
of M(atrix) theory as originally analyzed in \cite{Nair1}.

The same analysis is applicable to the case of $k=3$ and we can
show that the brane solution corresponding to $k=3$ also breaks
the supersymmetries. For the completion of discussion, we present
the factors $c_{ij\al}Q_\al \ga_{ij}$ for
$(i,j)=(9,10),(11,12),(13,14)$ in suitable choices of $c_{ij\al}$
and 6-dimensional gamma matrices:
\beqar
    c_{9\,10\al} \, Q_\al \, \ga_{9 \, 10}
    &\sim &
    \left( \sqrt{3} Q_3 + Q_8 + 2\sqrt{2} Q_{15} \right) \left(%
    \begin{array}{cccc}
      \si_1 & 0 & 0 & 0 \\
      0 & -\si_1 & 0 & 0 \\
      0 & 0 & \si_1 & 0 \\
      0 & 0 & 0 & -\si_1 \\
    \end{array}%
    \right) \nonumber\\
    c_{11\, 12 \al} \, Q_\al \, \ga_{11 \,12}
    &\sim &
    \left( - \sqrt{3} Q_3 +
    Q_8 + 2\sqrt{2} Q_{15} \right)\left(%
    \begin{array}{cccc}
      \si_1 & 0 & 0 & 0 \\
      0 & \si_1 & 0 & 0 \\
      0 & 0 & \si_1 & 0 \\
      0 & 0 & 0 & \si_1 \\
    \end{array}%
    \right) \label{s5}\\
    c_{13\,14\al} \, Q_\al \, \ga_{13 \,14}
    &\sim &
    \left(- 2 Q_8 +
    2\sqrt{2} Q_{15} \right)\left(%
    \begin{array}{cc}
      {\bf 1} & 0 \\
      0 & -{\bf 1} \\
    \end{array}%
    \right) \nonumber
\eeqar
where $Q_3$, $Q_8$ and $Q_{15}$ are the
$N^{(3)} \times N^{(3)}$ matrices representing diagonal parts of
$SU(4)$ algebra in the $(n,0)$-representation. In the last line,
${\bf 1}$ denotes the $4\times 4$ identity matrix.

%%%%%%%%%%%%%%%%%%%%%%%%%%%%%%%%%%%%%%%%%%%%%%%%%%%%%%%%%%%%%
\section{L7-branes and extra potentials in M(atrix) theory}

As we have seen in (\ref{c20}), the potential energy of a prospective
L7-brane with $\cp^3 \times S^1$ geometry is proportional to $n$,
leading to infinite energy in the large $N$ limit. In this section,
we introduce extra potentials to the bosonic part of the M(atrix) theory
Lagrangian so that the total potential energy of the L7-brane becomes
finite in the large $N$ limit.
From (\ref{c19}) we have found $\dot{r} \sim n^{-2}$. Thus the
kinetic energy of brane states with
$\cp^k \times S^1$ geometry is proportional to $\frac{N^{(k)}}{R} n^{-4}$.
Since the kinetic energy is suppressed by $n^{k-4}$, we
can consider the brane solution for any of $k=1,2,3$ as a static solution.
Consideration of potential energies will suffice for the
stability analysis of brane solutions.
In what follows, we first present a general form of the extra potentials
which is appropriate for our fuzzy $\cp^k$ brane solutions.
We then consider a few cases in detail, eventually obtaining a suitable form
of the extra potential for the emergence of L7-branes.

\vspace{0.4cm}
 \noindent \emph{4.1 \underline{Extra potentials: a general form}}
\vspace{0.2cm}

We consider the following form of potentials.
\beqar
    F_{2s+1} (X) &=& F_{[ij]^{s} \al }
    \Tr ( X_{i_1} X_{j_1}X_{i_2} X_{j_2}\cdots X_{i_r} X_{j_r} X_{\al} )
    \label{e1}\\
    F_{[ij]^{s} \al } &=&  \tr ( [ t_{i_1}, t_{j_1}] [ t_{i_2}, t_{j_2}]
    \cdots [ t_{i_s}, t_{j_s}] t_{\al} ) 
    \label{e2}
\eeqar
where $t_A$ ($A = i, \al$) are the generators of $SU(k+1)$ in the
fundamental representation with normalization $\tr ( t_A t_B ) = \hf \del_{AB}$.
As discussed earlier, $t_i$'s (including $t_j$'s) correspond to
the elements of $\underline{SU(k+1)} - \underline{U(k)}$ $(i = 1,2,\cdots , 2k)$
and $t_\al$ correspond to the elements of a $\underline{U(k)}$ subalgebra
$( \al = 1,2, \cdots , k^2$).
In the above expressions, $s$ takes the value of $s=1,2,\cdots, k$ and
$X_i$'s represent arbitrary matrix coordinates which
are, eventually, to be evaluated by the fuzzy $\cp^k$ coordinates $X_i = r(t) Q_i$.
Notice that the number of $X$'s is odd. This corresponds to the fact
that $F_{[ij]^{s} \al }$ are related to the rank-$(2s+1)$ invariant tensors of $SU(k+1)$.
We shall consider this point further in the next section.
In the following, we rather
show the correctness of the general form $F_{2s+1}$ in (\ref{e1}) for fuzzy $\cp^k$
brane solutions in M(atrix) theory.
The M(atrix) theory Lagrangian with the extra potential $F_{2s+1}$ is given
by
\beqar
    \L^{(2s+1)} &=& \L - \la_{2s+1}  F_{2s+1} (X) \label{e3}\\
    \L &=& \Tr \left( \frac{\dot{X_I}^2}{2R} + \frac{R}{4} [X_I , X_J]^2
    \right)
    \label{e4}
\eeqar
where $\L$ is the bosonic part of the original
M(atrix) theory Lagrangian (\ref{c1}) and $\la_{2s+1}$ is a coefficient of
the potential $F_{2s+1}$. The matrix equations of motion
are expressed as
\beq
    \frac{1}{R} {\ddot X}_I - R [ X_J , [X_I , X_J]] +
    \la_{2s+1} \frac{\del}{\del X_I} F_{2s+1} = 0 \, .
    \label{e5}
\eeq
Thus, in order to show the correctness of the general form in (\ref{e1}), it is
sufficient to see whether the term $\frac{\del}{\del X_I} F_{2s+1}$ is
proportional to the $Q_i$ when $X_I$ is evaluated by the ans\"{a}tze (\ref{c6}).

\vspace{0.4cm}
 \noindent \emph{4.2 \underline{Modification with $F_3$: Myers effect}}
\vspace{0.2cm}

For $s=1$, we have
\beq
    F_{[ij]\al} = \tr ( [ t_i , t_j ] t_\al ) = \frac{i}{2} c_{ij \al}
    \label{e6}
\eeq
where we use the normalization $\tr ( t_\al t_\bt ) = \hf \del_{\al \bt}$.
The potential $F_3 (X)$ is then written as
\beq
    F_3 (X) = \frac{i}{2} c_{ij \al} \Tr (X_{i} X_{j} X_\al ) \, .
    \label{e7}
\eeq
Since $c_{ij \al} \sim \ep_{ij \al}$,
the addition of $F_3$ to the M(atrix) theory Lagrangian essentially
leads to the so-called Myers effect from a viewpoint of IIA string theory \cite{Myers}.
Now we can calculate
\beqar
    \left. \frac{\del}{\del X_i} F_3 (X) \right|_{X=rQ} &=&
    \frac{i}{2} r^2 c_{ij \al} Q_j Q_\al \nonumber \\
    &=& \left( \frac{i r}{2} \right)^2  Q_i
    \label{e8}
\eeqar
where we use the relation (\ref{c10}).
Thus we find that the fuzzy $S^2$ remains the solution of M(atrix) theory
modified with the extra potential $F_3$.
As we shall see in a moment, generalizations along these lines can be
made for the potentials with higher ranks.

\vspace{0.4cm}
 \noindent \emph{4.3 \underline{Modification with $F_5$}}
\vspace{0.2cm}

For $s=2$, we have
\beqar
    F_{[ij]^{2} \al} &=& \tr ( [ t_{i_1} , t_{j_1} ][ t_{i_2} , t_{j_2} ] t_\al )
    \nonumber \\
    &=& i c_{i_{1}j_{1} \al_{1}} i c_{i_{2}j_{2} \al_{2}} \tr ( t_{\al_1}t_{\al_2}t_{\al} )
    \nonumber \\
    &=& - \frac{1}{4} c_{i_{1}j_{1} \al_{1}} c_{i_{2}j_{2} \al_{2}} d_{\al_{1} \al_{2} \al }
    \label{e9}
\eeqar
where we use the fact that $t_{\al_1}$ and $t_{\al_2}$ are commutative;
these generators correspond to ``diagonal'' elements of a $U(2)$ algebra
in terms of its matrix representation.
The symbol $d_{\al_{1}\al_{2}\al}$ is called the totally symmetric symbol of
$SU(k+1)$ and is defined by $d_{\al \bt \ga} = 2 \tr( \{ t_\al , t_\bt \} t_\ga )$.
The potential $F_5 (X)$ is then written as
\beq
    F_5 (X) =
    - \frac{1}{4} c_{i_{1}j_{1} \al_{1}} c_{i_{2}j_{2} \al_{2}} d_{\al_{1} \al_{2} \al }
    \Tr ( X_{i_1} X_{j_1} X_{i_2} X_{j_2} X_{\al} ) \, .
    \label{e10}
\eeq
This is a natural generalization of the Myers term (\ref{e7}) to a higher rank.
Notice that $F_{5}$ exists for any $SU(k+1)$ with $k \ge 2$.
The variation of $F_5$ with respect to $X_{i_1}$ is expressed as
\beqar
    \left. \frac{\del}{\del X_{i_1}} F_5 (X) \right|_{X=rQ} &=&
    - \frac{1}{4} r^4 c_{i_{1}j_{1} \al_{1}} c_{i_{2}j_{2} \al_{2}} d_{\al_{1} \al_{2} \al }
    Q_{j_1}
    \underbrace{Q_{i_2}Q_{j_2}}_{\frac{i}{2} \frac{c_{i_{2}j_{2}\bt_{2}}}{\sqrt{C_{2}^{(k)}}} Q_{\bt_2}}
    Q_\al \nonumber \\
    &=& \left( \frac{i}{2} \right)^3 \frac{r^4}{\sqrt{C_{2}^{(k)}}} c_{i_{1}j_{1} \al_{1}}
    \underbrace{d_{ \al_{1} \al_{2} \al } Q_{j_1} Q_{\al_2} Q_\al }_{ c_{k,n} Q_{j_1} Q_{\al_1}}
    \nonumber \\
    &=& \left( \frac{i r}{2} \right)^4 \frac{c_{k,n}}{C_{2}^{(k)}} Q_{i_1}
    \label{e11}
\eeqar
where we evaluate the variation with the fuzzy $\cp^k$ ans\"{a}tze (\ref{c6}),
using the relations (\ref{a3}), (\ref{c7}) and (\ref{c10}).
The result (\ref{e11}) shows that the fuzzy $\cp^k$ ($k=2,3$) remain the solutions
of M(atrix) theory even if it is modified with the extra potential $F_5 (X)$.

In this case, the matrix equations of motion (\ref{e5}) become
\beq
    \left[  \frac{\ddot{r}}{R} +
    \frac{R}{C^{(k)}_{2}} r^3 \left( 1 + \frac{\la_5 r}{16 R} c_{k,n} \right) \right] Q_{i}
    = 0 \, .
    \label{e12}
\eeq
This matrix equation is then reduced to an equation of $r(t)$ as in the
case of the pure bosonic M(atrix) theory.
We can easily carry out the evaluation of $F_5$ on the fuzzy $\cp^k$ ans\"{a}tze as
\beqar
    F_5 (rQ) &=& - \frac{1}{4} r^5
    c_{i_{1}j_{1} \al_{1}} c_{i_{2}j_{2} \al_{2}} d_{\al_{1} \al_{2} \al }
    \Tr ( Q_{i_1} Q_{j_1} Q_{i_2} Q_{j_2}Q_{\al} ) \nonumber\\
    &=& \left( \frac{i}{2} \right)^{4} \frac{r^5}{C^{(k)}_{2}}
    \underbrace{
    d_{ \al_1 \al_2 \al} \Tr ( Q_{\al_1} Q_{\al_2} Q_{\al} )
    }_{ c_{k,n} \Tr (Q_{\al} Q_{\al} )} \nonumber \\
    &=&
    \frac{k^2}{k^2 + 2k} \frac{r^5 c_{k,n}}{16 C^{(k)}_{2}} N^{(k)} ~ \sim ~ n^{k-2}  r^5
    \label{e13}
\eeqar
where we use the relation (\ref{c14}).
Notice that the $n$ dependence of (\ref{e13}) is the same as
that of the M(atrix) theory potential in (\ref{c20}).

\vspace{0.4cm}
 \noindent \emph{4.4 \underline{Modification with $F_7$}}
\vspace{0.2cm}

Since $s \le k$ and we are interested in $k=1,2,3$,
the case of $s=3$ is allowed only for $k=3$. In this case, we have
\beqar
    F_{[ij]^{3} \al}
    &=& \tr ( [ t_{i_1} , t_{j_1} ][ t_{i_2} , t_{j_2} ][ t_{i_3} , t_{j_3} ] t_\al )
    \nonumber \\
    &=& - i c_{i_{1}j_{1} \al_{1}}  c_{i_{2}j_{2} \al_{2}} c_{i_{3}j_{3} \al_{3}}
    \tr ( t_{\al_1} t_{\al_2} t_{\al_3} t_{\al} )
    \label{e14}
\eeqar
where, as in the case of $F_5$, $t_\al$'s
are corresponding to ``diagonal'' generators of $U(3)$.
Thus they are commutative to each other.
Anticommutation relations of these are given by
\beq
    \{t_{\al} , t_{\bt} \} = d_{\al \bt \ga} t_\ga
    \label{e15}
\eeq
where the symmetric symbol $d_{\al \bt \ga}$ is that of $SU(k+1)$
but its indices refer only to a $U(3)$ subgroup.
Notice that a $U(1)$ element is included in this subgroup;
for $SU(4)$ (corresponding to $k=3$) the $U(1)$ element is
conventionally chosen by $t_{15}$ and this choice would be
used for any $SU(k+1)$ ($k \ge 3$).
Using (\ref{e15}), we then find
\beqar
    F_{[ij]^{3} \al} &=& - \frac{i}{4}
    c_{i_{1}j_{1} \al_{1}}  c_{i_{2}j_{2} \al_{2}} c_{i_{3}j_{3} \al_{3}}
    d_{\al_1 \al_2 \bt} d_{\bt \al_3 \al } \, ,
    \label{e16}\\
    F_7 (X) &=& F_{[ij]^{3} \al}  \Tr ( X_{i_1} X_{j_1} X_{i_2} X_{j_2} X_{i_3} X_{j_3} X_{\al} ) \, .
    \label{e17}
\eeqar
The variation of $F_7$ with respect to $X_{i_1}$ is then expressed as
\beqar
    \left. \frac{\del}{\del X_{i_1}} F_7 (X) \right|_{X=rQ} &=&
    - \frac{i}{4} r^6 c_{i_{1}j_{1} \al_{1}}  c_{i_{2}j_{2} \al_{2}} c_{i_{3}j_{3} \al_{3}}
    d_{\al_1 \al_2 \bt} d_{\bt \al_3 \al }
    Q_{j_1}
    \underbrace{Q_{i_2}Q_{j_2}}_{\frac{i}{2} \frac{c_{i_{2}j_{2}\bt_{2}}}{\sqrt{C_{2}^{(k)}}} Q_{\bt_2}}
    \underbrace{Q_{i_3}Q_{j_3}}_{\frac{i}{2} \frac{c_{i_{3}j_{3}\bt_{3}}}{\sqrt{C_{2}^{(k)}}} Q_{\bt_3}}
    Q_\al \nonumber \\
    &=& i \left( \frac{i}{2} \right)^4 \frac{r^6}{C_{2}^{(k)}} c_{i_{1}j_{1} \al_{1}}
    d_{\al_1 \al_2 \bt} d_{\bt \al_3 \al } Q_{j_1} Q_{\al_2} Q_{\al_3} Q_\al
    \nonumber \\
    &=& i \left( \frac{i}{2} \right)^4 \frac{r^6}{C_{2}^{(k)}} c_{i_{1}j_{1} \al_{1}}
    c_{k,n}^{~2} Q_{j_1} Q_{\al_1}
    \nonumber \\
    &=&
    \left( \frac{i r}{2} \right)^6 \frac{2 c_{k,n}^{~2}}{\sqrt{C_{2}^{(k)}}^3 } Q_{i_1}
    \label{e18}
\eeqar
where we use the relation (\ref{a3}), {\it i.e.},
$d_{ \al \bt \ga } Q_{\al}Q_{\bt}  = c_{k,n} Q_{\ga}$, twice.
Notice that the symmetric symbol $d_{\al \bt i}$ vanishes as discussed in (\ref{c11-1}).
The result (\ref{e18}) shows that the fuzzy $\cp^k$ ($k=3$) remains as a solution
to M(atrix) theory even if it is modified with the extra potential $F_7 (X)$.
Lastly we can evaluate $F_7$ on the fuzzy $\cp^k$ ans\"{a}tze as
\beqar
    F_7 (rQ) &=&
    - \frac{i}{4} r^6 c_{i_{1}j_{1} \al_{1}}  c_{i_{2}j_{2} \al_{2}} c_{i_{3}j_{3} \al_{3}}
    d_{\al_1 \al_2 \bt} d_{\bt \al_3 \al }
    \Tr (Q_{i_1}Q_{j_1}Q_{i_2}Q_{j_2}Q_{i_3}Q_{j_3}Q_{\al} )
    \nonumber \\
    &=&
    i \left( \frac{i}{2} \right)^5 \frac{r^7}{\sqrt{C_{2}^{(k)}}^3 }
    \underbrace{d_{\al_1 \al_2 \bt} d_{\bt \al_3 \al }
    \Tr (Q_{\al_1}Q_{\al_2}Q_{\al_3}Q_{\al} )}_{ c_{k,n}^{~2} \Tr (Q_{\bt}Q_{\bt})}
    \nonumber \\
    &=& - \frac{k^2}{k^2 + 2k } \frac{r^7 c_{k,n}^{~2}}{32 \sqrt{C_{2}^{(k)}}^3 } N^{(k)}
    \, \sim \, n^{k-3} r^7 \, .
    \label{e19}
\eeqar

\vspace{0.4cm}
 \noindent \emph{4.5 \underline{Emergence of L7-branes}}
\vspace{0.2cm}

To recapitulate, we are allowed to include the extra potentials of the
form $F_{2s+1} (X)$ ($s \le k$, $k=1,2,3$) in the M(atrix) theory
Lagrangian as far as the brane solutions of $\cp^k$ geometry in the
transverse directions are concerned. Evaluated on the fuzzy $\cp^k$ ans\"{a}tze,
these extra potentials are expressed as
\beqar
    F_3 (rQ) &=& -\frac{k^2}{k^2 + 2k } \frac{r^3}{4 \sqrt{C_{2}^{(k)}}} N^{(k)}
    \, \sim \, n^{k-1} r^3
    \label{l1} \\
    F_5 (rQ) &=& \frac{k^2}{k^2 + 2k } \frac{r^5 c_{k,n}}{16 C_{2}^{(k)}} N^{(k)}
    \, \sim \, n^{k-2} r^5
    \label{l2} \\
    F_7 (rQ) &=& - \frac{k^2}{k^2 + 2k } \frac{r^7 c_{k,n}^{~2}}{32 \sqrt{C_{2}^{(k)}}^3 } N^{(k)}
    \, \sim \, n^{k-3} r^7
    \label{l3} \\
    V (rQ) &=& \frac{k^2}{k^2 + 2k }\frac{R r^4}{4 C_{2}^{(k)}} N^{(k)}
    \, \sim \, n^{k-2} R r^4
    \label{l4}
\eeqar
where we include the M(atrix) theory potential in (\ref{c20}).
As mentioned earlier, we consider a static solution. Thus the effective
Lagrangian for the static solution is given by
\beq
    \L_{eff} = - V_{tot} (r) = - V (rQ) - \la_3 F_3 (rQ) -
    \la_5 F_5 (rQ) - \la_7 F_7 (rQ) \, .
    \label{l5}
\eeq
From (\ref{l1})-(\ref{l4}), we can express $V_{tot} (r)$ as
\beqar
    V_{tot}(r) &=& \frac{k^2}{k^2 + 2k }\frac{R}{C_{2}^{(k)}} N^{(k)} v(r)
    ~ \sim ~ n^{k-2} R
    \label{l6}\\
    v(r) &=& \frac{r^4}{4} - \mu_3 r^3 + \mu_5 r^5 + \mu_7 r^7
    \label{l7}
\eeqar
where
\beq
    \mu_3 = \frac{\la_3}{4 R} \sqrt{C_{2}^{(k)}} \, , ~~
    \mu_5 = \frac{\la_5}{16 R} c_{k,n} \, , ~~
    \mu_7 = - \frac{\la_7}{32 R} \frac{c_{k,n}^{~2}}{\sqrt{C_{2}^{(k)}}}  \, .
    \label{l8}
\eeq

In the case of $k=1$, only $F_3$ exists and the potential $v(r)$ becomes
$v_3 (r) \equiv \frac{r^4}{4} - \mu_3 r^3$.
This potential is relevant to the Myers effect. In Myers' analysis
\cite{Myers}, the coefficient $\la_3$ is determined such that it
satisfies the equations of motion $\frac{\d v_3}{\d r} = r^3 - 3 \mu_3
r^2 = 0$.
Thus we have $\mu_3 \sim r/3 \sim 1$ ($r > 0$), or $\la_3 \sim R/n$.
Analogously, we may require $\la_5 \sim R$, $\la_7 \sim n R$ such that
$v(r) \sim 1$.
Notice that we demand $\mu_{5}, \mu_{7} > 0$
so that the potential $v(r)$ is bounded below; otherwise
the solutions become unphysical in the limit of large $r$.
We also demand  $\mu_3 > 0$ such that $v(r)$ always has a minimum at $r>0$;
regarding the range of $r$, we require $r > 0$ because
it describes a size of each brain solution.

The total potential $V_{tot} (\sim n^{k-2} R)$ becomes finite for $k=1,2$
in the large $n$ limit.
In this limit, the brane solutions corresponding to $k=1,2$ therefore exist
regardless the value of $v(r)$.
For $k=3$, however, $V_{tot}(r)$ diverges in the
large $n$ limit {\it unless} $v(r) = 0$.

To further investigate the case of $k=3$, we now consider the following
potential $v_7 (r)\equiv \frac{r^4}{4} - \mu_3 r^3 + \mu_7 r^7 $, without
a $F_5$ term.
The equation of motion for $r$ is given by
\beq
    \frac{\d v_7}{\d r} = 7 \mu_7 r^2
    \left( r^4 + \frac{r}{7 \mu_7} - \frac{3 \mu_3}{7 \mu_7} \right) = 0 \, .
    \label{l9}
\eeq
Denoting the nonzero solution by $r_*$, we now plug
this back to $v_7 (r)$; $v_7 (r_*) = \frac{r_{*}^{3}}{7} (
\frac{3r_*}{4} - 4 \mu_3 )$. If we fix $\mu_3$ as $\mu_3 = \frac{3}{16}
r_*$, $v_7(r_*)$ vanishes. In this case, $V_{tot} (r_*)$ becomes finite
in the large $n$ limit and the corresponding L7-branes are allowed to
present as a stable solution at the minimum $r=r_*$. The L7-branes exist
for a particular value of $\mu_3$. In this sense, the strength of the
$F_3$ flux can be considered as a controlling parameter for the
emergence of L7-branes. The same analysis applies to a potential without
$F_7$; $v_5 (r) \equiv  \frac{r^4}{4} - \mu_3 r^3 + \mu_5 r^5$. If we
consider the full potential $v(r)$ with nonzero $\mu_{2s+1}$ ($s=1,2,3$),
the existence of L7-branes can be similarly shown at the minimum of $v(r)$,
with two of the three $\mu_{2s+1}$ serving as the controlling
parameters.

There are few remarks on the existence of the L7-brane solutions.
Firstly, if we introduce fluctuations from the minima, the potential $v(r)$
becomes nonzero and consequently the total potential $V_{tot}(r)$
diverges in the large $n$ limit. In other words, fluctuations from the
stabilized L7-branes are suppressed.
Secondly, the involving extra potentials are expressed as
$F_{2s+1} (rQ) \sim  \Tr {\bf 1}$
where ${\bf 1}$ is the $N^{(3)} \times N^{(3)}$ identity matrix.
These can be regarded as constant matrix-valued potentials.
This fact suggests that the analysis in the previous section also holds
with $F_{2s+1}(rQ)$, preserving the L7-brane solutions non-supersymmetric.
Lastly, in terms of M(atrix) theory as a 11-dimensional theory,
the emergence of L7-branes and the suppression of their
fluctuations suggest a compactification of the theory down to 7 dimensions.
We shall discuss this point further in the next section.

%%%%%%%%%%%%%%%%%%%%%%%%%%%%%%%%%%%%%%%%%%%%%%%%%%%%%%%%%%%%%%%%%%%
\section{Compactification scenarios in M(atrix) theory}

As mentioned in the introduction, the existence of a 7-form suggests a
compactification of the 11-dimensional theory down to 7 or 4 dimensions.
In this section, we first show that the extra potential $F_7 (X)$ in (\ref{e14})
can be considered as a 7-form in M(atrix) theory.
We then discuss that the effective
Lagrangian (\ref{l5}) with $k=3$ can be used
for a compactification model of M(atrix) theory down to 7 dimensions.
We also consider a compactification scenario of M(atrix)theory
down to 4 dimensions by use of fuzzy $S^4$ which can be defined in
terms of fuzzy $\cp^3$ \cite{Abe1}.

\vspace{0.4cm}
 \noindent \emph{5.1 \underline{$F_{2s+1}$ as matrix differential forms: a cohomology analysis}}
\vspace{0.2cm}

The general expression of $F_{2s+1}(X)$ in (\ref{e1}) is closely related
to differential $(2s+1)$-forms of $SU(k+1)$ ($s=1,2,\cdots , k$).
Differential forms of $SU(k+1)$ are in general constructed by the Lie algebra
valued one-form
\beq
    g^{-1}d g = -i t_{A} E_{A}^{a} d \th^a =-i t_{A} E_A
    \label{d1}
\eeq
where $g=\exp(-i t^a \th^a)$ is an element of
$SU(k+1)$, $\th^a$'s are continuous group parameters, $t_A$'s are
generators of $SU(k+1)$ in the fundamental representation with normalization
$\tr(t_A t_B)=\hf \del_{AB}$, and $E_A = E_{A}^{a} (\th) d \th^a$ are
one-form frame fields on $SU(k+1)$ ($a, A=1,2,\cdots, k^2 + 2k$).
The differential $(2s+1)$-forms $\Om^{(2s+1)}$ of $SU(k+1)$ are then
defined as
\beqar
    \Om^{(2s+1)} &=& \tr (g^{-1} d g)^{2s+1} \nonumber\\
    &=& (-i)^{2s+1} \tr(t_{A_1}t_{A_2}\cdots t_{A_{2s+1}})E_{A_1}
    \wedge E_{A_2} \wedge \cdots \wedge E_{A_{2s+1}}  \nonumber\\
    &=& F_{A_1 A_2 \cdots A_{2s+1}} E_{A_1} \wedge E_{A_2} \wedge \cdots \wedge E_{A_{2s+1}} \, ,
    \label{d2} \\
    F_{A_1 A_2 \cdots A_{2s+1}} &=& \frac{(-i)^{2s+1}}{2^s}
    \tr ( [ t_{A_1} , t_{A_2} ] [ t_{A_3} , t_{A_4} ]  \cdots [ t_{A_{2s-1}} , t_{A_{2s}} ]
    t_{A_{2s+1}} ) \, .
    \label{d3}
\eeqar
Notice that the invariant tensor $F_{A_1 A_2 \cdots A_{2s+1}}$
is essentially the same as the tensor $F_{[ij]^{s} \al}$
defined in (\ref{e2}).
The only difference, apart from proportionality coefficients,
is the index assignments.
A peculiar form in $F_{[ij]^{s} \al}$
arises from the fact that we are interested in algebraic properties
of $\cp^{k} = SU(k+1)/U(k)$ rather than the full $SU(k+1)$.
In other words, $F_{[ij]^{s} \al}$ is a subset of
the invariant tensor $F_{A_1 A_2 \cdots A_{2s+1}}$.
The possible number of such tensors is $k$ $( \ge s)$;
these tensors are called the Casimir invariants for the Lie group $SU(k+1)$.

Mathematically, it is known that
the differential $(2s+1)$-forms $\Om^{(2s+1)}$ of $SU(k+1)$
are elements of $\H^{2s+1} (SU(k+1), {\bf R})$, {\it i.e.},
the $(2s+1)$-th cohomology group of $SU(k+1)$ ($s=1,2,\cdots, k$) over the real numbers.
The Casimir invariants $F_{A_1 A_2 \cdots A_{2s+1}}$
are in one-to-one correspondence with cohomology classes for the Lie
group $SU(k+1)$. This correspondence is related to the so-called Weil
homomorphism between Casimir invariants and Chern classes.
For descriptions of these mathematical aspects of $\Om^{(2s+1)}$, one may refer to \cite{Nair3}.

From the above argument, we can interpret the potentials $F_{2s+1}(X)$ in (\ref{e1})
as matrix-valued differential forms, or as fuzzification of the
differential forms $\Om^{(2s+1)}$ in (\ref{d2}); the fuzzification may be
carried out by replacing $E_A$ with arbitrary matrices $X_A$.
In the following, we justify this statement by showing cohomology properties of
$F_{2s+1}(X)$ evaluated on fuzzy $\cp^k$.
In other words, we shall see that
$F_{2s+1}(X)$, evaluated on fuzzy $\cp^k$, can be considered as
matrix-valued forms that are closed but not exact.

As we have shown in (\ref{e8}), (\ref{e11}) and (\ref{e18}), variations of
$F_{2s+1} (X)$ $(s=1,2,3)$ with respect to $X_i$ are linear in $Q_i$
when $X$'s are evaluated on on the fuzzy $\cp^k$ ans\"{a}tze  $X_i =r(t)Q_i$.
Since $Q_i$ are traceless matrices, this corresponds to the fact that
$F_{(2s+1)}(rQ)$ are matrix-valued {\it closed} differential forms.

On the other hand, as shown in (\ref{l1})-(\ref{l3}),
$F_{2s+1} (rQ)$ $(s=1,2,3)$ are nonzero constants.
This arises from the fact that
$F_{(2s+1)}(rQ)$ are matrix-valued {\it non-exact} differential forms.
Notice that the non-exactness of an ordinary differential form, say $\Om^{(3)}$, can
be shown by $\int_{S^3} \Om^{(3)} \ne 0$, where the integration is taken
over $SU(2) = S^3$.
(If $\Om^{(3)}$ is exact, {\it i.e.}, $\Om^{(3)} = d \al$,
Stokes' theorem says $ \int_{S^3} \Om^{(3)} = \int_{\d S^3} \al$ where
$\d S^3$ is the boundary of $S^3$. Since $S^3$ is a compact manifold, $
\int_{\d S^3} \al = 0$. Thus $\Om^{(3)}$ can not be exact. One can
similarly show the non-exactness of $\Om^{(2s+1)}$ in general, using the
fact that the volume element of $SU(k+1)$ can be constructed in terms of
the wedge products of $\Om^{(2s+1)}$'s.)
$F_3 (Q)$ is a fuzzy analogue of $ \int_{S^3} \Om^{(3)}$.
Thus the value of $F_3 (Q)$ in (\ref{l1})
corresponds to the nonzero volume element of a fuzzy version of $S^3$.
Locally, we may parametrize $S^3$ as $S^3 \approx \cp^1 \times S^1$.
Thus $F_3 (rQ)$ can also be seen as the volume element of a fuzzy version of
$\cp^1 \times S^1$.
Analogously, we can make a local argument to show
that $F_{2k+1} (rQ)$ ($k=2,3$) correspond to the volume elements of fuzzy
versions of $S^{2k+1} \approx \cp^k \times S^1$.
(Note that since $\cp^k
= S^{2k+1}/ S^1$, we can locally express $S^{2k+1}$ as $\cp^k \times
S^1$ in general.)
We can therefore interpret $F_{2s+1} (rQ)$ as
matrix versions or fuzzifications of $(2s+1)$-forms $\Om^{(2s+1)}$,
given that the invariant tensors $F_{A_1 A_2 \cdots A_{2s+1}}$ in (\ref{d3})
are restricted to the form of $F_{[ij]^{s} \al}$ defined in (\ref{e2}).

\vspace{0.4cm}
 \noindent \emph{5.2 \underline{Freund-Rubin type compactification}}
\vspace{0.2cm}

The fact that we can interpret $F_{(7)}(rQ)$ as a 7-form
in M(atrix) theory is interesting in search for a compactification model
of M(atrix) theory. As mentioned in the introduction, according to
Freund and Rubin \cite{FR}, existence of a differential $d^{\prime}$-form in
$d$-dimensional theories suggests compactification of $(d-d^{\prime})$ or $d^{\prime}$
space-like dimensions ($d^{\prime} < d$). Usually the Freund-Rubin type
compactification is considered in 11-dimensional supergravity which
contains a 4-form. Although this compactification has a problem in
regard to the existence of chiral fermions, the Freund-Rubin
compactification of M-theory has been shown to avoid such a problem and
presumably provides a realistic model of M-theory in lower dimensions
\cite{Acha}. The presence of the above-mentioned 7-form then supports a
possibility of the Freund-Rubin type compactification in M(atrix) theory.
It is not clear at this point how the effective Lagrangian
(\ref{l5}) relates to compactified 7-dimensional supergravity in the low energy limit.
However, as discussed before, the Lagrangian (\ref{l5})
with $k=3$ does capture a desirable physical property for the
compactification of M(atrix) theory down to 7 dimensions.

In terms of the 11-dimensional M-theory, the potential $F_7 (rQ)$
corresponds to a flux on a curved space of $( \cp^3 \times S^1 )\times
\M_4$ geometry where $\M_4$ is some four-dimensional manifold.
The Freund-Rubin type compactification requires that the
manifold $\M_4$ be a positively curved Einstein manifold.
This suggests that we in fact have to describe $\M_4$
by some fuzzy spaces, say, fuzzy $\cp^2$ or fuzzy $S^4$ in the context of M(atrix) theory.
So far we have neglected the contributions from $\M_4$ in the
fuzzy $\cp^k$ brane solutions (\ref{c6}) where we squash irrelevant directions.
We can however include $\M_4$ contributions to the M(atrix) theory potential (\ref{l6})
such that they do not affect the existence condition for the L7-branes, namely, the
finiteness of $V_{tot} (r)$ in the large $n$ limit.
Notice that there is freedom to add an $n$-independent constant
to $V_{tot}(r)$.
Such a case is possible, for example, if we identify $\M_4$
with a relatively small-size fuzzy $S^4$.

It is known that fuzzy $S^4$ can be represented by block-diagonal matrices, with
their full matrix dimensions given by $N^{(3)}$ \cite{Abe1}.
Thus it is natural to parametrize $\M_4$ by fuzzy $S^4$
for $n$-independent modifications of the Lagrangian (\ref{l5}) with $k=3$.
Notice that one of the four dimensions in $\M_4$ represents
the time component in M(atrix) theory.
Thus a naive application of fuzzy $S^4$ to the geometry of $\M_4$
is not suitable for the framework of M(atrix) theory.
However, as in the case of the IKKT model \cite{IKKT}, one can consider
the time component in terms of a matrix.
As far as a matrix model building of M-theory in the large
$N$ limit is concerned, we may then parametrize $\M_4$ in terms of fuzzy $S^4$.
Along the line of these considerations, we can therefore
interpret the Lagrangian (\ref{l5}) with $k=3$ as an effective
Lagrangian for a compactification model of M(atrix) theory down to 7 dimensions.

\vspace{0.4cm}
 \noindent \emph{5.3 \underline{Construction of fuzzy $S^4$: a brief review}}
\vspace{0.2cm}

Compactification of M(atrix) theory down to 4 dimensions is also
possible for the Freund-Rubin compactification in the presence of the 7-form.
In what follows, we shall discuss this possibility by use of fuzzy $S^4$.
For this purpose, we first give a brief review of the construction of fuzzy $S^4$.
It is known that functions on fuzzy $S^4$
can be constructed from functions on fuzzy $\cp^3$ by imposing the following
constraint \cite{Abe1}:
\beq
    [ \F( Q_i ) , Q_{\tilde{\al}} ] = 0
    \label{d4}
\eeq
where $\F( Q_i )$ are arbitrary polynomial
functions of the fuzzy $\cp^3$ coordinates $Q_i$
($i=1,2,\cdots, 6$ or, in a conventional choice of $SU(4)$ generators,
$i=9,10,\cdots, 14$).
The index $\tilde{\al}$ corresponds to the algebra of
$\tilde{H} = SU(2) \times U(1)$ in terms of the decomposition of $SU(4)
\rightarrow SU(2)\times SU(2) \times U(1)$.
In this decomposition, two $SU(2)$'s and one $U(1)$ are defined by
\beq
    \left(%
    \begin{array}{cc}
    \underline{SU(2)} & 0 \\
    0 & 0 \\
    \end{array}%
    \right)  , ~
    \left(%
    \begin{array}{cc}
    0 & 0 \\
    0 & \underline{SU(2)} \\
    \end{array}%
    \right) , ~
    \left(%
    \begin{array}{cc}
    1 & 0 \\
    0 & -1 \\
    \end{array}%
    \right)
    \label{d5}
\eeq
in terms of the ($4 \times 4$)-matrix
generators of $SU(4)$ in the fundamental representation.
In the above expressions, $\underline{SU(2)}$ denotes the algebra of the $SU(2)$ group in the
($2\times 2$)-matrix representation and $1$ represents the ($2\times 2$) identity matrix.
With an imposition of (\ref{d4}), the functions on fuzzy $\cp^3$, $\F( Q_i )$,
are reduced to functions on fuzzy $S^4$.

As analyzed in \cite{Abe1}, upon the imposition of (\ref{d4}) the fuzzy
$\cp^3$ coordinates $Q_i$ become fuzzy $S^4$ coordinates, say, $Y_\mu$
($\mu=1,2,3,4$). These are no longer represented by full
$N^{(3)} \times N^{(3)}$ matrices but by
$N^{(3)} \times N^{(3)}$ {\it block-diagonal} matrices.
The block-diagonal matrix $Y_\mu$ is composed of $(n+2-m)$ blocks of dimension $m$
for $m=1,2,\cdots, n+1$ and can be expressed as
\beq
    Y_\mu = \mbox{block-diag} ( \underbrace{1,1,\cdots, 1}_{n+1},
    \underbrace{\Box_2, \Box_2,\cdots, \Box_2}_{n}, \cdots,
    \Box_{n},\Box_n , \Box_{n+1} )
    \label{d6}
\eeq
where $\Box_{m}$ denotes a full $(m \times m)$ block matrix.
Notice that the matrix dimension of $Y_\mu$ remains as
\beq
    \sum_{m=1}^{n+1} (n+2-m)m = \frac{1}{6}(n+1)(n+2)(n+3)=N^{(3)} \, ,
    \label{d6-1}
\eeq
while the number of nonzero matrix elements becomes
\beq
    \sum_{m=1}^{n+1} (n+2-m) m^2 = \frac{1}{12}(n+1)(n+2)^2 (n+3) \equiv N^{S^4} \, .
    \label{d6-2}
\eeq
We can in fact show that the number $N^{S^4}$ corresponds to
the number of coefficients in a mode expansion of
truncated functions on $S^4$.
(For details of the correspondence between
fuzzy $S^4$ and truncated functions on $S^4$, see \cite{Abe1}.)
From the expression (\ref{d6}), we can easily tell that
$Y_\mu$ commute with $N^{(1)} \times N^{(1)}$ block
matrices where $N^{(1)}=n+1$ is the number of 1's in (\ref{d6}).
Furthermore, $Q_{\tilde{\al}}$ is in an $N^{(1)} \times N^{(1)}$
matrix representation of $\underline{SU(2)}$ in terms of the
decomposition of $SU(4)$ discussed in (\ref{d5}).
Thus, from the expression (\ref{d6}), we can check that
$Y_\mu$ indeed satisfies the constraint (\ref{d4}).

The configuration (\ref{d6}) may be the most
natural one in comparison with fuzzy $\cp^3$ but it is not the only one
that describes fuzzy $S^4$.
For example, we can locate the same-size blocks in a single block,
following some operation, say, matrix multiplication or matrix addition,
instead of diagonally locating each block one by one.
The dimension of the alternative matrix configuration is then given by
\beq
    \sum_{m=1}^{n+1} m = \hf (n+1)(n+2) = N^{(2)} \, .
    \label{d6-3}
\eeq
This means that fuzzy $S^4$ can also be described by
$N^{(2)} \times N^{(2)}$ block-diagonal matrices, say, $\tilde{Y}_\mu$.

\vspace{0.4cm}
 \noindent \emph{5.4 \underline{Emergence of fuzzy $S^4$}}
\vspace{0.2cm}

We now consider an imposition of the constraint (\ref{d4})
on the effective Lagrangian (\ref{l5}) with $k=3$.
Since the potentials $F_{2s+1}(rQ)$ are proportional to the identity matrix,
they are not affected by the constraint (\ref{d4}) and
the local coordinates of fuzzy $\cp^3$ $Q_i$ are simply replaced by the
fuzzy $S^4$ coordinates $Y_\mu$ after the imposition of (\ref{d4}).
Corresponding matrix equations of motion become linear in $Y_\mu$.
Thus, as in the case of the L7-brane solutions, we can similarly
consider emergence of L5-branes with fuzzy $S^4$ geometry
as brane solutions to modified M(atrix) theories.
As before, the emergence of such L5-branes can be argued by requiring
that the potential energy of the branes at minima of the
total potential energy becomes finite.

In terms of the local coordinates of fuzzy $\cp^3$ $Q_i$, the M(atrix)
theory potential is calculated as
$\Tr \frac{Rr^4}{4} [ Q_i , Q_j]^2 = -
\frac{N^{(3)}}{15} \frac{Rr^4}{ C_{2}^{(3)}}$.
The sum of the extra potentials for the emergence of L7-branes has been
given by $\frac{N^{(3)}}{15} \frac{R r_{*}^{4}}{ C_{2}^{(3)}}$ where
$r_*$ represent a minimum of $v(r)$ in (\ref{l7}).
In terms of the local coordinates of fuzzy $S^4$ $Y_\mu$,
a matrix Lagrangian for the emergence of the spherical L5-branes is then expressed as
\beq
    \L_{S^4 \times S^1} = \Tr \left( \frac{\dot{r}^{2} Y_{\mu}^{2}}{2R}
    + \frac{R r^4}{4} [Y_\mu , Y_\nu]^2 + \frac{R r_{*}^{4}} {15 C^{(3)}_{2}}
    {\bf 1}_{N^{(3)}} \right)
    \label{d7}
\eeq
where we include the kinetic term which is zero for static solutions.
The value of $r_*$ is determined by the controlling parameters for the emergence
of the spherical L5-branes. For example, consider the potential $v(r)$
of the form $v_5 (r) = \frac{r^4}{4} - \mu_3 r^3 + \mu_5 r^5$ where
$\mu_3$, $\mu_5$ are given by (\ref{l8}) with $k=3$. In this case, the
controlling parameter is given by $\mu_3$ as discussed before. From
$\left. \frac{\d v_5}{\d r} \right|_{r_*} =0$ and $v_5 (r_*) =0$, we can
easily find $r_* = 8 \mu_3$. Notice that $r_*$ is independent of $n$
since $\mu_3$ is an $n$-independent parameter.

In order to obtain compactification of M(atrix) theory down to 4
dimensions, we simply eliminate the longitudinal direction in the spherical L5-branes.
The relevant brane solution would be a transverse 4-brane of fuzzy $S^4$ geometry.
Apparently, this brane solution does not have a time component
in the framework of M(atrix) theory but, as mentioned
earlier, it is possible to express the time component by a matrix
as far as a matrix model building of M-theory in the large $N$ limit is concerned.
Bearing this possibility in mind, we can conjecture an action for
such a fuzzy $S^4$ solution as
\beqar
    \S_4 & = & \frac{r^4 R}{4} \, \Tr \left( [Y_\mu , Y_\nu ]^2 ~+~
    \frac{\bt}{C^{(3)}_{2}} \, {\bf 1}_{N^{(3)}} \right) \, ,
    \label{d8} \\
    \bt &=& \frac{4}{15} \left( \frac{r_*}{r} \right)^4 \, \sim \, 1 \, .
    \label{d9}
\eeqar
There are basically two fundamental parameters, $R$ and $N=N^{(3)} \sim n^3$.
We consider that in the large $N/R$ limit the matrix action (\ref{d8})
describes compactification of M-theory in 4 dimensions.
$R$ is essentially the 11-dimensional Planck length $l_p$; remember that
$R$ is given by $R = gl_s = g^{2/3} l_p$ where $g$ is the
string coupling constant and $l_s$ is the string length scale.
There are no restrictions on the size parameter $r$. This suggests conformal invariance
of the theory of interest. The parameter $\bt$, on the other hand, will be determined
by how we carry out flux compactifications in terms of controlling parameters.
Since the fuzzy $S^4$ solutions are constructed from the L7-branes of $\cp^3 \times S^1$ geometry
on top of the algebraic constraint (\ref{d4}), these solutions are also non-supersymmetric.
Lastly we would like to emphasize that the above action can be used as
a physically interesting 4-dimensional matrix model of M-theory compactification.

\vspace{0.4cm}
 \noindent \emph{5.5 \underline{Purely spherical L5-branes as new solutions in M(atrix) theory}}
\vspace{0.2cm}

As we have discussed in (\ref{d6-3}), fuzzy $S^4$ can also be represented by
$N^{(2)}\times N^{(2)}$ block-diagonal matrices $\tilde{Y}_\mu$.
Its matrix dimension is the same as that of fuzzy $\cp^2$.
Thus, as in the case of fuzzy $\cp^2$ solutions,
there are no problems on infinite energy and
we can obtain an L5-brane of $S^4\times S^1$ geometry as a solution to
the original M(atrix) theory without any extra potentials.

The transverse directions of this L5-brane are {\it purely} spherical.
Notice that it is different from the previously proposed spherical L5-brane \cite{CLT}.
The previous solution has been constructed under a condition \cite{CLT}:
\beq
    \ep_{ijklm} X_i X_j X_k X_{l} \sim X_{m}
    \label{d10}
\eeq
where $X_i$'s ($i=1,2,\cdots , 5$) denote matrix coordinates of the
brane solution, four out of five coordinates representing the transverse directions.
Owing to the Levi-Civita tensor, the above condition makes sense when
indices $i, j, \cdots , m$ are distinctive one another.
Strictly speaking, the transverse directions following the condition (\ref{d10})
do not describe $S^4$ geometry but rather part of $\cp^3$ geometry.
In the context of fuzzy $\cp^3$ solutions developed in the present paper,
this can easily be seen by rewriting the above condition as
\beq
    c_{ij \al} c_{kl \bt}d_{\al \bt \ga} Q_i Q_j Q_k Q_l \sim
    d_{\al \bt \ga} Q_\al Q_\bt \sim Q_\ga
    \label{d11}
\eeq
where we replace $\ep_{ijklm}$ by $c_{ij \al} c_{kl \bt}d_{\al \bt \ga}$
and $X_i$'s by the fuzzy $\cp^3$ coordinates $Q_i$.
As we have seen in (\ref{e9}), $c_{ij \al} c_{kl \bt}d_{\al \bt \ga}$
corresponds to the rank-five invariant tensor of $SU(4)$.
Explicit proportionality in (\ref{d11}) can be read from (\ref{e11}).

As discussed above, in order to obtain purely spherical geometry, we
need to impose an algebraic constraint on $Q_i$.
The resultant solution then becomes an L5-brane of fuzzy $S^4$ geometry
in the transverse directions, Fluctuations of this brane solution can naturally
be described by $Q_i \rightarrow  Q_i + A_i$.
As mentioned in the introduction, there has been a difficulty to include
fluctuations in the previously proposed spherical L5-branes \cite{CLT}.
Our version of a purely spherical L5-brane avoids this difficulty
and provides a new brane solution to M(atrix) theory.

%%%%%%%%%%%%%%%%%%%%%%%%%%%%%%%%%%%%%%%%%%%%%%%%%%%%%%%%%%%%%%%
\section{Conclusions}

In the present paper, some of the previously known brane solutions in M(atrix) theory are
reviewed in a systematic manner by use of the fuzzy complex projective
spaces $\cp^k$ ($k=1,2,\cdots$) as ans\"{a}tze for the solutions.
We show that this particular type of ans\"{a}tze for $k \le 4$ indeed
satisfies the M(atrix) theory equations of motion.
For the cases of $k \le 3$, we have checked that the brane solutions break all
supersymmetries of M(atrix) theory.
An L7-brane solution corresponding to $k=3$ has an infinite potential energy
in the large $N$ limit.
We can however make it finite and can show the existence of static
L7-brane solutions with an introduction of extra potentials.
Such potentials, which are
closely related to differential $(2r+1)$-forms of $SU(k+1)$ ($r=1,2,\cdots, k$),
can be simplified to the identity matrices up to some constants.
We show that even with these potentials the fuzzy $\cp^k$ ($k \le 3$)
remain solutions to modified M(atrix) theories, possessing finite
potential energy in the large $N$ limit.
In the case of $k = 3$, this means that fluctuations from the L7-brane solution are suppressed
in the large $N$ limit and that we have a peculiar compactification scenario of
M(atrix) theory down to 7 dimensions.
This model can be analyzed by the effective Lagrangian given in (\ref{l5}) with $k=3$.

In the context of Freund-Rubin type compactification of M-theory, the
very existence of the 7-form implies compactification of the theory down to 7
or 4 spacetime dimensions.
This suggests that our analysis can be used to
give a physically interesting compactification to 4 dimensions.
As an example of such possibility, we have conjectured a compactified model of
M(atrix) theory in 4 dimensions, utilizing the definition of fuzzy $S^4$ in terms of fuzzy $\cp^3$.
The resultant action (\ref{d8}) is
expressed in terms of the coordinates of fuzzy $S^4$ (\ref{d6}).
Along the way, we also find the existence of new L5-branes in
M(atrix) theory which have purely spherical geometry in the transverse directions.

%%%%%%%%%%%%%%%%%%%%%%%%%%%%%%%%%%%%
\vskip .2in
\noindent {\bf Acknowledgments}
\vskip .06in\noindent
The author would like to thank Professor V.P. Nair for helpful discussions.
The author is also grateful to the referees of this paper for critical and valuable comments.

%%%%%%%%%%%%%%%%%%%%%%%%%%%%%%%%%%%%%%%%%%%%%%%%%%%%%%%%%%%%%%%%

\end{document}